\documentclass[12pt]{article}
\usepackage{graphicx}
\usepackage{latexsym}

\begin{document}

\title{On observation of neutron quantum states in Earth gravitational field at Laue-Langevin Institute, Grenoble}
\author{Anatoli Andrei Vankov\\         
% Declares the author's name. Optional title:
{\small \it IPPE, Obninsk, Russia; Bethany College, KS, USA,  anatolivankov@hotmail.com}}
%,\\\
%{\small \it 16 Shawmut Avenue, Mansfield, MA 02048, USA}}

\date{}

\maketitle

\begin{abstract}

L'expérience exécutée avec les neutrons ultra-froids à l'Institut de Laue-Langevin, Grenoble, est analysée en vue de la revendication que "les états quantiques à neutrons dans le champ gravitationnel de Terre" sont observés. Notre conclusion est que la susdite revendication n'est ni théoriquement ni expérimentalement justifiée. Nous critiquons aussi la déclaration que "l'observation des états quantiques gravitationnellement reliés de neutrons et des techniques expérimentales liées fournit un instrument unique à une large gamme d'enquêtes dans la physique fondamentale de particules et de champs".

\medskip
The experiment performed with ultra-cold neutrons at the Laue-Langevin Institute, Grenoble, is analyzed in view of the claim that ``neutron quantum states in Earth gravitational field'' are observed. Our conclusion is that the above claim is neither theoretically nor experimentally substantiated. We also criticize the statement that ``observation of the gravitationally bound quantum states of neutrons and the related experimental techniques provide a unique tool for a broad range of investigations in fundamental physics of particles and fields''. 

\medskip
 Key words: quantum mechanics; quantum gravity; ultracold neutrons; neutron quantum states; Earth gravitational field; experiment; Laue-Langevin Institute; Grenoble.

\
 
\end{abstract}
%\maketitle

%\section{Experiment set-up, objectives, and claims}
%\section{Experiment description}

\section{Introduction}

One of Physics Frontiers hot issues is quantum gravity and crave for either a new, break-through quantum gravity theory or General Relativity extension to the quantum domain. Different approaches to this problem did not succeed yet, \cite{Rovelli}, \cite{Smolin}, and elsewhere. The situation is aggravated by the fact that phenomena that could reveal any quantum properties of gravitational interaction or force transmitting mechanism have not been observed. At the same time, it is known from  Quantum Mechanics (QM) that bound quantum states of test particle can exist in a potential well formed by any force, the gravitational one being not excluded. A minimal (quantum) energy of neutron-Earth interaction on land is assessed about $10^{-12}$  $eV$. Therefore, ultracold neutrons (UCNs) can form, in principle, quantum layers, or ``quantum bouncer'' states \cite{Ignatovich}. 

Since, approximately, 1999, an experimental study of quantum bouncer phenomenon has been conducted at the Laue-Langevin Institute, Grenoble. In a series of publications, [4-17] (new), [18-25] (preceding), the authors describe methodology of the experiment and obtained results. Here, we refer to collections of works by {\em the authors of the Grenoble experiment} (unfortunately, we could not find a single publication devoted to final experimental results).

In brief, the main purpose of the Grenoble experiment is to study quantum states of UCNs confined in a horizontally placed neutron waveguide (further called  ``the slit'') in the gravitational field of Earth. The problem formulation and the experiment methodology are outlined within the QM theory with an addition of semi-classical ``phenomenology''. The experimental facility includes the neutron source (nuclear thermal reactor), the slit (a neutron waveguide), and an external detector (a counter of neutrons passed through the slit). The slit has a length $\Delta x$ about 15 $cm$ and a front rectangular aperture with a vertical width $a=\Delta z$ between bottom and upper parallel reflectors/mirrors. The width is variable in the range of about 2 - 60 
$mk$m. A wider horizontal width $\Delta y$ is fixed. 

There is a collimating UCN guide system to transport neutrons from the reactor to the slit. They enter the slit through the front aperture (window) at the speed uniformly spread in the range of about 4-10 $m/s$. Velocity vectors have some angular distribution. Upon reflections from the bottom mirror, neutron start bouncing and keep propagating in the horizontal direction. Due to grazing incidence, neutrons acquire a small vertical momentum component, which is also randomly distributed. This is the component which is subjected to quantization. However, how exactly the quantum states are formed and what should be practically measured to observe the effect is far from being evident. 

As a matter of fact, the authors decided to count transmitted through the slit UCNs at different widths. An external integral counter of transmitted neutrons at some distance from the exit is used in most of the measurements but sometimes the so-called position sensitive detector (PSD) is used as well. The PSD is a thin film that contains fission nuclei and is placed instead of the integral detector. As ``a differential detector'', the PSD seems to be advantageous for the experiment purpose because it gives a differential $z$-distribution of neutron flux. In any way, it is not clear how the counts of transmitted neutrons are related to the eigenvalue problem and the corresponding quantization picture expected by the authors. 

The authors claim that ``the quantum bouncer'' phenomenon has been observed in the experiment. In this connection, there were critical reactions to the experiment in publications where some technical and theoretical questions were raised but with no factual analysis  \cite{Critic}.  Interested physicists who are familiar with the experiment seem to be left with a belief that the neutron quantum states in the gravitational field have been, indeed, observed. The reason for this paper is that we do not share this belief.

The present work is devoted to the analysis of the experiment. We conclude that the published results do not support the statement on ``quantum bouncer'' observation, because, in our view, there are methodological flaws in experiment setup and measured data treatment. We also pay attention to the authors' work motivation that, as seen from publications, goes quite beyond the QM realm, as next: ``observation of the gravitationally bound quantum states of neutrons and the related experimental techniques provide a unique tool for a broad range of investigations in fundamental physics of particles and fields''  \cite{Nesvizh2}. We want to make it clear that ``the quantum bouncer'' has nothing to do with ``quantum gravity'' (a theory not formulated yet), or any quantum field theory concepts. The Grenoble experiment is the first attempt to practically realize the passive (non-radiating) QM system, therefore, it is academically interesting but in no way aimed at Physics Frontiers. Specific authors' claims concerning fundamental values of the experiment and its methodology are discussed later in more details.

In our analysis of the experiment, we proceed through two stages: 1. the academic formulation of the problem; 2. the real experiment.

\section{QM concept of confined particle}

\subsection{Examples of infinite well and quantum bouncer as opposed to field-driven QM systems}

For rather a pedagogical purpose, let us start with issues related to the general QM problem of confinement and the quantum bouncer, in particular. Examples of QM eigenvalue problem of particle in a potential well are well known from textbooks in the context of the 1D time independent Schrodinger equation
\begin{equation}
-\frac{\hbar^2}{2m}\frac{d^2 u}{dz^2}  + W(z) u(z)=E u(z)
\label{1}
\end{equation}
A popular illustration is a particle confined in the infinite well. The main point there is that the well must be suitable for standing wave formation so that a momentum of particle under the corresponding boundary conditions is quantized, in accordance with the de Broglie wave concept. Consequently, other quantities related to the momentum are characterized by discrete numbers.

Dealing with duality of the quantum object image, one can describe ``quantization'' in the infinite well in  
terms of {\em particles} in a motional quantum states (unlike in classical mechanics, a confined particle in QM cannot be at rest). Thus, a discrete set of particle energies is obtained $E_n=h^2 n^2 /2 m a^2$. In terms of {\em waves}, one should associate quantum energies to fundamental frequencies $f_n=E_n/h$ subjected to boundary conditions when the width $a$ fits an integer number of wavelength. Basically, the quantization is a direct consequence of the requirement that a wave amplitude is zero at the boundaries (the particle is confined). 

Actually, one has to admit both even and odd types of solutions to (\ref{1}) by requiring that not the wavelength but half wavelength fits the well. Then, $n$ should be replaced by half integer $n/2$, and one finds twice as many levels:
$2a= n \lambda = n h/(2m E)^{1/2}\  =>  \ E_n=(nh)^2/8 m a^2$
where the total energy is solely the kinetic one. For the fixed $a$, the set $E_n$ has to be found. Of course, one can fix the energy value $E$ to find the corresponding width set $a_n$. 

The solution in the above example constitutes a set of normalized harmonic functions \ $u_n(z)$. \ A set of wave functions is similarly obtained in the eigenvalue problem for a potential well of any shape. The complete set is useful for Fourier  series decompositions of any function describing ``mixed states'', which can arise due to random experimental conditions, a geometry perturbation, and other causes. In any case, a stationary ``mixed'' state can be approximated by a linear combination of base eigenfunctions (eigenvectors) and the corresponding eigenvalues (fundamental frequencies). The full study of such a system includes issues of stability of modes, wave packet spreading, energy conservation etc, in both Schrodinger and Heisenberg time dependent representations. 

One can notice commonality between a particle in the infinite well and a neutron in the slit (the gravitational bouncer): both belong to a type of QM systems conceptually disconnected from a field theory. They are ``tuned'' to resonant modes by means of geometrical configuration but cannot be ``excited'' by external radiation. In this sense, they may be termed ``passive systems''. Clearly, the gravitational quantum bouncer presents a strictly nonradiating  ``Earth plus neutron plus mirror'' macroscopic system, which exploits an electric property of material (elasticity), on the one hand, and an attraction of particle by Earth, on the other hand. The quantization concept in this case is merely a demonstration of de Broglie waves under certain (kinematical) boundary conditions.

Field-driven systems are principally different. They include existing man-made and natural quantum systems of dynamically interacting particles forming collective standing waves synchronously with fundamental proper frequencies.  Excitation states (metastable resonance modes) get populated as a result of action of an external radiation interacting with the system by means of field quanta exchange. Such dynamically ``active'' systems are described by well established field theories and routinely studied in physical laboratories. A laser pumping technique and an atomic oscillator spectrometry are examples of QED practical application. 

Next, we focus our attention on the quantum bouncer concept.

\subsection{Confinement and Airy equation}

The equation (\ref{1}) for a general form of potential well $W(z)$ can be rewritten in terms of variable wavenumber $k(z)$
\begin{equation} 
\psi'' + k^2 (z) \psi(z)=0, \ \ p(z)=\hbar k(z)=\pm\sqrt{2m[E-W(z)]}
\label{2}
\end{equation}
The confinement implies that $W(z)$ monotonously grows as $|z|\to \infty$ so that $E>W(z)$ inside the well and  $E<W(z)$ beyond the boundary outlined by pairs of return points where $E_n=W(z_n)$. The momentum (the wavenumber) takes imaginary values in the confinement region and real values outside the well. The eigenvalue problem usually requires a junction of solutions for $E<W(z)$ and $E>W(z)$. In other words, the ``exact'' (numerical) solution and the ``modified'' one due to specific boundary conditions (quantization constraints) are not the same thing.

The usual general approach to the problem is to express a wave function inside the well in the form  
$\psi=A\exp{[(\imath / \hbar)S]}$ for real $A(z)$ and $S(z)$. The substitution yields to the equations:
\  $S'^2=p^2+\hbar^2 (A''/A)$ \  and \   $S'' A + 2S' A'=0$,\ solutions of which are sensitive to boundary conditions in regions of left and right return points. The ``best'' modified solution is achieved for high frequency and/or slowly changing amplitude, when terms of order $\hbar^2$ can be ignored (the quasi-classical WKB approximation). 

In the linear potential model, the quantity $F=-W(z)/z$ plays a formal role of parameter of a boundary shape of the potential well, and the acceleration $g=F/m$ is not necessarily ``recognized'' as a field strength. Moreover, the Equivalence Principle and conservative properties of the field have no impact on the Schrodinger problem formulation, unless the properties are brought into the solution. One can elaborate another example, where a charged particle (say, a proton) serves as a quantum bouncer in a static macroscopic  electric field. Obviously, a realization of such an idea would add nothing to our current knowledge of electricity. Figuratively speaking, ``Schrodinger does not distinguish between, say, Newton and Coulomb''. 

There is an ample literature related to different aspects of the linear potential problem \cite{QM}. For a  linear attractive potential (physically not specified yet) $W(z)=Fz$, $F=-mg$, the Schrodinger equation takes the second order differential form $\psi''-\xi\psi=0$, called the Airy equation \cite{AiryEq}. In this case, the Airy solution  $Ai(z)$, we are interested in, is shown in Fig. \ref{Airy} (notice a change of units:\  
$z\to \xi$, $E\to \epsilon$), $s=\xi -\epsilon$). There are 7 nodes shown there, correspondingly up to 
$z=-z_{7}$ in the $z$ scale. 

The equation reflects the next confinement problem: for the potential well formed by the bottom mirror $W \to\infty$ at $z=-z_N$ and the boundary line $W(z)=Fz$, find the set of eigenvalues $E_N=F |z_n|$,  $n=1, 2, ... N$, ($N$ fixed) inside the well $E_n\ge W(z)$. For $n=1$ we have the quantum bouncer in the  state of lowest energy. Here the coordinate system is chosen such that $W(0)=0$, so one seeks for a solution in the interval $-z_N\le z\le 0$. A neutron is supposed to be dropped from rest at the classical return point $z=0$ and to accelerate in free fall in the negative direction of the axis $z$ within the width of the slit $a=|z_N|$. It is seen that the Airy solution does not satisfy the quantization requirement $Ai(0)=0$. The difficulty (already noted) arising in the neighborhood of the classical return point $z_0=0$, where $p(z)\to 0$ as $z\to z_0$, is supposed to be mended by the solution modification (WKB approximation, in particular).

%%%%%%%%%%%%%%%%%%%%%%%%%%%%%%%%
\begin{figure}[t]
\includegraphics{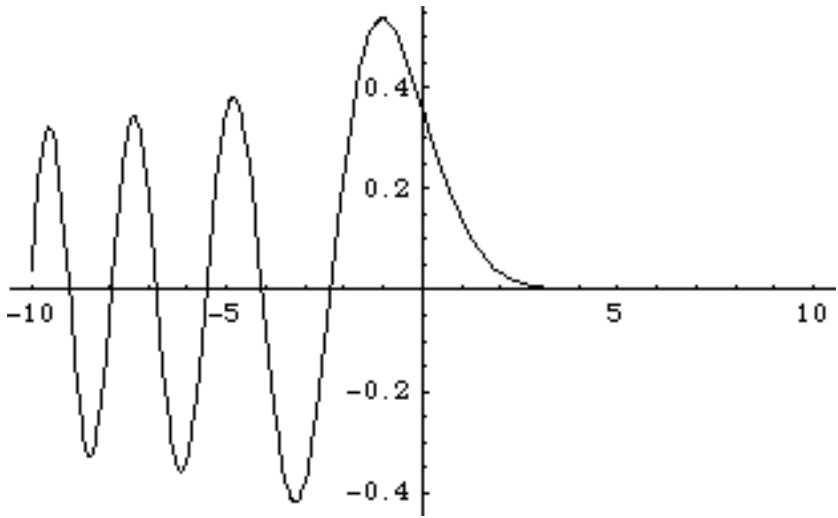}
\label{Airy}
\caption{\label{Airy}  The Airy function $Ai(s)$ for linear potential energy.  $s=\xi-\epsilon $, \ $\xi=(z/\tilde z_0)^{3/2}$,\  $\epsilon =E/mg\tilde z_0$,\  $\tilde z_0=(\hbar/2m^2 g)^{1/3}$. 
$Ai(s)\to 0$ as $s\to\pm\infty $}
\end{figure}
%%%%%%%%%%%%%%%%%%%%%%%%%%%%%%%
%While real experiment conditions will be discussed later, here some comments in the academically formulated context seem to be useful. The first comment 

There are two issues here. The first one concerns the confinement concept and the requirement that the classical return point must be fixed (for example, at $z=0$) to specify the boundary and quantization conditions. A physical meaning is as next. One has to release a neutron from a tiny source attached to some non-absorbing frame, such as a mirror, at $z=0$. For a neutron in projectile motion, one can also consider drop points at random within the width. In this case, the solution should be presented in the form of superposition of base functions with coefficients to be determined as a function of width. In the authors' work, the notion of ``classical return point'' is different: it is considered in the context of Airy functions ``penetrating'' into classically forbidden space. We argue, however, that the tail in the region $z\ge 0$ is not the quantum tunneling effect but arises because the problem is ill posed and its solution needs to be modified, as discussed.

The second issue relates to the total energy conservation law in the gravitational field. Classical laws (the Newton gravitational law, in particular) reside in average characteristics of quantum assemble (the Ehrenfest theorem). In our example of fixed drop point (say, $z=0$), the total energy conservation law comes to the scene,  $E=K(z) + W(z)=const$ ($const=0$ is chosen) so that neutrons hitting the bottom mirror have the same total energy regardless of distance passed. A constant value cannot be ``quantized''. To formulate the eigenvalue problem consistently with the energy conservation, the initial conditions should be redefined (e. g. \cite{Prentis} and elsewhere). It could be done by fixing a position of the bottom mirror $z_N=0$ (instead of the upper one) to allow neutrons to be dropped, one by one, from distances $z$ at random above the mirror. 

Recall, the Airy functions $Ai_n (z)$ in Fig. \ref{Airy} are originally defined in the coordinate system where nodes $i=1, 2, ... N$ are counted from the fixed classical return (drop) point $z=0$ in the direction of momentum increase (or distance $|z|$ passed). This is a negative direction of $z$ axis. With the origin of the $z$-axis at the bottom mirror, the Airy functions are defined in the range $z\ge 0$. This means that, for a statistical assemble of neutrons, a multitude of drop points randomly distributed within the width must be considered in calculations of the resulting probability density $P_{sum}(z)$ in a new coordinate system. The latter must be found by a change of variable  $z => z'(z, n)$ for each individual function $Ai_n(z)$. 

Now each neutron in a statistical assemble will be found in the mixed state, a superposition of eigenfunctions defined in the half-space $z\ge 0$, with coefficients depending on the width and a drop point coordinate. The total energy is constant in time for each neutron (the energy conservation law). 

The authors seem to take this scheme without discussions. It is illustrated in Fig. \ref{quantBouncer}   \cite{Nesvizh15} in the example of squared Airy functions $|Ai_n(z)|^2$,\ $(z>0) $. However, the drop points are not random there but chosen at node points $z_n> 0$  for $n=1,\ 2,\ 3,\ 4$ (in WKB terms). Notice, the functions are not properly normalized: their maximal amplitudes are taken equal. The picture illustrates contributions of four first states in the resulting probability density $P_{sum}(z)$ for a width of the slit $a=z_N$  ($N=4$ in the picture). Here, with the $z$-coordinate increase, a numeration of nodes $z_i$ of individual functions $|Ai_n (z)|^2$ must go backward (compare with Fig. \ref{Airy}): for $n=4$, $i=4, 3, 2, 1$; for $n=3$, $i=3, 2, 1$; for $n=2$, $i=2, 1$, and for $n=1$, $i=1$. The energy $E_n=mgz_n$ rises with $n= 1,\ 2,\ ... N$. 

By summing up the graphs, one will get a picture of resulting neutron probability density $P_{sum}(z)$. Obviously,  the quantum pattern becomes dispersed due to overlapping of wavefunctions of different numbers and shifting of nodes for different modes. Here, this happened when drop points were chosen not randomly but at node points satisfying the quantization conditions. The quantum pattern will be even more smooth for drop points  randomly distributed. 

The important lesson drawn from this subsection is, as next.  A single random neutron in the slit is in a mixed state characterized by a superposition of ``pure states'' with coefficients monotonously dependent on the width and the drop point coordinate. In a statistical neutron assemble, probability density functions overlap so that the resulting density  $P_{sum}(z, a)$ as a function of width $a$ is visualized as a smooth curve. This is the first sign of trouble with ``neutron quantum state observation''.

We shall see that from the authors' phenomenological model another prediction follows: a distinct quantum pattern (in the form of peaks). We disagree with that. More new details
will come with the discussions of the real experiment.
%%%%%%%%%%%%%%%%%%%%%%%%%%%%%
\begin{figure}[t]
\includegraphics{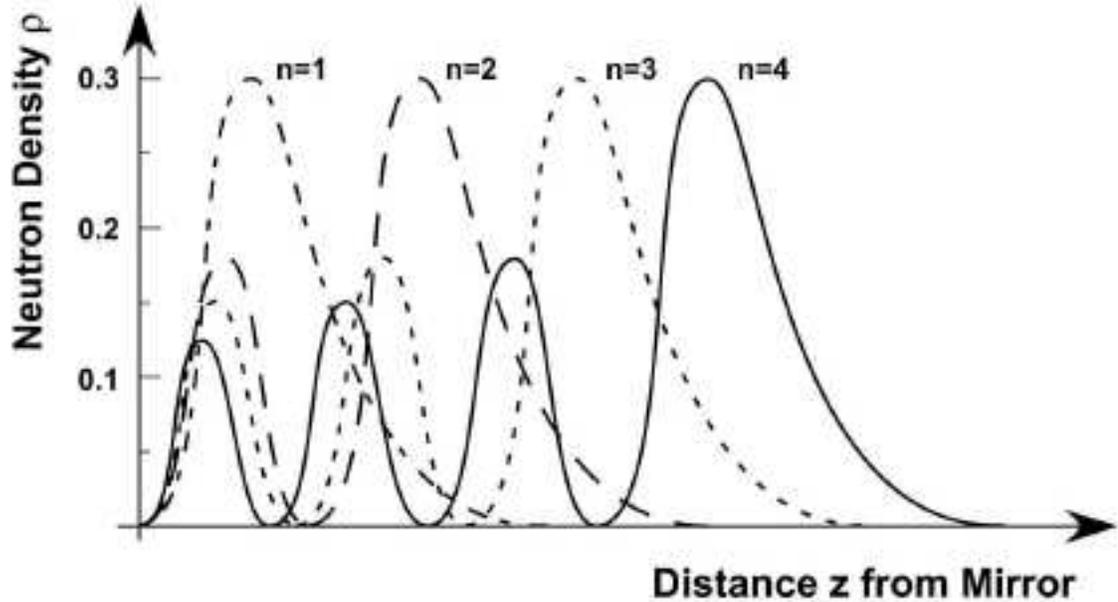}
\label{quantBouncer}
\caption{\label{quantBouncer} 
Probability density from Airy solutions for four first states $E_n$: $n=1, 2, 3, 4$; bottom mirror position is fixed at  $z=0$; neutrons dropped from distances $z=z_n$ in the upper half-space $0 \le z \le a$; $a$ \- \ width of slit. By summing up the graphs, one gets a picture of resulting probability density $P_{sum}(z)$ \cite{Nesvizh15}.}
\end{figure}
%%%%%%%%%%%%%%%%%%%%%%%

\subsection{Classical versus quantum probability density}

In connection with Fig. \ref{quantBouncer}, it is interesting to consider a classical probability density.  There is a standard method of its derivation but the authors devised, for no particular reason, their own method based on a consideration of ``phase space'' (not discussed here). 

For a neutron dropped from the point $z=a$ above the mirror, the probability of finding a particle within the small interval $(z, z+\Delta z)$ is defined: 
$P_{clas}(z) dz=C (\Delta t/\Delta z) dz = C dz/v(z)$, where the normalization constant is $C=1/\int_0^a P(z) dz$, and  $v(z)=\sqrt{2g(a-z)}$. Finally:
\begin{equation}  
P_{clas}(z) dz= dz/2 \sqrt{a(a-z)} ,\  \  \ (0\le z\le a)
\label{3}
\end{equation}

As seen from (\ref{3}), the form of classical density distribution for a single drop does not depend explicitly on $g$, as it should be for a constant acceleration. In this connection, it would be instructive to consider 
the problems of the infinite well and the slit in a semi-classical model in the case when neutrons are dropped from a fixed point. 

Consider a slit of width $a=z_1$, where $z_1=\lambda_1 /2$ is the de Broglie half-wavelength of a neutron dropped from rest. The neutron travels a distance $z=p^2/2mg$, where $p=h/z$. From this:
\begin{equation}
z_1=(h^2/8 m^2 g)^{1/3}; \  \  \   E_1 (sl)=mgz_1=mv^2 /2=hf_1 (sl)
\label{4}
\end{equation}
%\begin{equation}
% E_1 (sl)=mgz_1=mv^2/2=hf_1(sl)
%\label{5}
%\end{equation}
where $v$ is a final speed, $f_1 (sl)=1/\Delta t (sl)$ is the frequency and $\Delta t (sl)$ the corresponding time interval of a neutron oscillation,  $\Delta t_1 (sl)=z_1/ \bar v$, and $\bar v=v/2$ is the average speed. In this example, the average speed but not the acceleration $g$ matters.  

One can find that the neutron lowest level in the slit coincides with that in the infinite well of width $a=z_1$. Recall, for the infinite well  \ $E_1 (i.w.)=h^2/2m a^2 $. It is straightforward to check the next: if $a=z_1$, then $E_1 (sl)=E_1 (i.w.)$ and vice versa. The same is true if the width is chosen equal to any number of half-wavelengths. Thus, a measurement within just one lowest energy interval is not indicative of the role of gravitational field in the slit experiment. 

%Next we discuss the experimental results and their interpretation by the authors. After that, we shall continue the methodological analysis, first of all, the methods of neutron injection and quantum state observation.

\section{Grenoble experiment realities}

\subsection{Methodology of the experiment, and ``the principle for observation of quantum gravitational states''}

\subsubsection{Neutron injection method and projectile motion}

So far, we considered an academic formulation of the problem. A situation in the real experiment 2D rime-dependent conditions is severely aggravated by the method of injection, when neutrons are launched into a projectile motion at a high speed, specifically, in the range 4 {\--} 10 $m/s$ that is, in the energy range of about $(1 {\--} 5)\cdot 10^{-7}$\ $eV$. It should be compared with the speed of about few $cm/s$ in vertical ``quantum'' oscillation and the corresponding energies of quantum states of order $10^{-12}$\ $eV$. Only a fraction of about  $10^{-5}$ of actual kinetic energy ${\tilde E}$ of neutrons in the slit is subjected to quantization in question. Neutrons of actual energies are able to climb a potential hill (on land) up to several meters, while ``quantization'' is supposed to take place within a distance about $(1 {\--} 6)\ 10^{-5}$\ $m$.  How the quantum spectral characteristics of a tiny fraction of quantized momentum/energy of neutrons on the huge background of fluctuating total energy could be resolved and analyzed in an experimental apparatus is a big question not clearly explained in the authors' works.

One of the ``damaging'' effects was previously discussed. It comes from the fact that trajectories are characterized by a continuous (random) distribution of drop point coordinates $z_{\alpha}$. A single neutron should be found in a ``mixed'' state described by a superposition of base functions in the properly chosen coordinate system 
 $\Psi_{\alpha}(z) \sim \sum_n {C_n(z_{\alpha, a}) \psi_n (z)}$,\ \ $(n=1, 2,  ...N)$, where $N$ is limited by the width $a$. A monotonous dependence of the coefficients  $C_n$ on the width $a$ (for a given $z_{\alpha}$) leads to overlapping of the functions (the ``smoothening'' effect in quantum pattern. After averaging over 
 a random distribution of $z_{\alpha}$ within the slit, the smoothing effect will be even grater so that the resulting density distribution must be a monotonous function of $a$: 
\begin{equation}
  P_{sum}(z, a) \sim   \int {d/dz_{\alpha} [P_{\alpha}(z_{\alpha}\to z)]dz_{\alpha}},\ \ \ (0<z<z_{\alpha}<a)  
\label{5}
\end{equation}
Thus, we have the methodological problem of a fuzzy image of the quantum bouncer due to its statistical nature. This is a part of the primary (incoherent neutron source) problem.

Another part of the problem is due to neutron scattering in the slit that also causes a great ``damaging'' effect: a huge background of ``unwanted neutrons'', which acquire random vertical components of the momentum during reflections from the bottom mirror. Consider, for example, a neutron with a horizontal speed  $v_0$=4 \ $m/s$. As a result of specular reflection at a small angle of inclination about $\theta =3.5\cdot 10^{-3}$, the neutron acquires a velocity vertical component about $v_z$ =1.4 \ $cm/s$, that is a quantum bouncer characteristic. 

To remove ``unwanted'' neutrons, the authors put a foreign body into the quantum system, an absorber. As a result, only those quantum bouncers that ``jumping'' vertically appreciably lower than the absorber height are left in the slit. However, the absorber does not eliminate the problem of fuzzy quantum image. Moreover, it creates a new problem.  The absorber works as a neutron sink. At small heights, it kills neutrons in the ground state (the quantum bouncer). Indeed, a bouncing (vertical) amplitude fluctuates in accordance with the Heisenberg uncertainty. For small widths  $a\le z_1$, the uncertainty principle fully works: 
 $\Delta z_1 \Delta p_z \sim \Delta t_1 \Delta E_1  \sim h$, where $\Delta E_1 =mg z_1$;\  $z_1=\bar {v_1} \Delta t_1$.  In the course of bouncing, neutrons eventually hit the absorber, what results in a blockage of their transmission. The blockage means a cut-off in the transmission curve. For a greater absorber heights, only neutrons of energy $E<E_{n+1}$ have a chance to survive in the slit of width $a\approx z_n$,\ $n\ge 2$.

The cut-off could be pronounced due to additional factors:  a) reflections off the slit of incident neutrons
diffracted on the front aperture ; b) absence of neutrons of small speed $v< 4$\ $m/s$ in the incident neutron spectrum. After the cut-off at about  $z_1$, a transmission curve must be a monotonously increasing function. However, there could be some `irregularities'' in the curve due to residual effects of diffraction of neutrons on the exit window.

To illustrate the problem, consider a neutron having an absorption probability $p(z)$ per one bounce at some absorber height $z$. On average, a quantum bouncer makes about 15 bounces in the slit. Hence, a chance to pass through the slit is $F(z)=[1-x(z)]^{15}$; for example, for $x=0.2$,\ it is about $F= 0.035$. This is how a sharp cut-off is produced.

The measurements, indeed, show a smooth enough curve with a cut-off at a small height. This tells us that there is a serious methodological problem of quantum state formation and observation. Our comments on how the authors manage it are next.

\subsubsection{``The principle of observation of quantum states''}

To interprete the measurements, the authors introduced ``the principle of observation of quantum states'' based on the authors' phenomenological model. The remarkable feature of ``the principle'' is that it literarily cuts the Gordian knot of all bothering us problems. Basically, it consists of two statements:

1. A distinctly observable quantum pattern of neutrons in the slit is formed (this is what the phenomenological model says).

2. PSD counts of transmitted through the slit neutrons at different absorber heights adequately reproduce the differential quantum probability density function. Similar counts by the integral detector (the transmission function) adequately reproduce the corresponding integral quantum probability density function. In accordance with the phenomenological model, the transmission function has a stepwise form that translates into the corresponding distinct quantum pattern from the phenomenological model. Namely, it is understood that the probability density distribution has the form of resonance lines $\delta(z-z_n)\delta(E-mgz_n)$.

In accordance with the principle of observation, the plan of the experiment was to count passing through the slit neutrons by a position sensitive detector (PSD) at some distance from the back window. In practice, an integral flux of passing through the slit neutrons was most of the time measured as a function of width (``the transmission function'' $F(a)$). The method of counts normalization is not clear from publications. The following interpretation of the transmission function is given \cite{Nesvizh6}, \cite{Nesvizh7}: 

{\em ``Below about 15 $\mu m$, no neutrons can pass the slit... Ideally, we expect a stepwise dependence of transmission as a function of width. If the width is smaller than the spatial width of the lowest quantum state, then transmission will be zero. When the width is equal to the spatial width of the lowest quantum state, the transmission will increase sharply. A further increase in the width should not increase the transmission as long as the width is smaller than the spatial width of the second quantum state. Then again, the transmission should increase stepwise. At sufficiently high slit width one approaches the classical dependence... It was found, that except for the ground state, the stepwise increase is mostly washed out.''} 

Here, the integral transmission $F(z)$ as a function of width is meant that in some way must resemble the integral $z$-distribution of the probability density $\int_0^z {P(z') dz'}$. In the phenomenological model, neutrons in the slit are characterized in field theory terms such as a state lifetime, a state transition probability, etc.  (\cite{Nesvizh2}, \cite{Nesvizh18}, and other authors' works).  See the next extract: 

{\em `` Significant increase of the accuracy of experiments of this kind could be achieved by the long storage of neutrons in quantum states and by a measurement of resonant transitions between them, thus allowing one directly to calculate the energies of the corresponding quantum states''.}

%Read also about a notion of state superposition:

%{ \em ``In fact, the transversal motion of neutrons in the wave-guide can be described as a superposition of the neutron wave-guide transversal modes:}
%\     $\Phi (z, t)=\sum_n {C_n \Psi_n (z)\exp{(-\imath E_n t-\Gamma_n t/ 2 )}}$ ''\

%Here, $\Gamma$ is proportional to the absorption probability, the registered flux of transmitted neutrons is 
%\    $F=\int_0^\infty {|\Psi(z, \tau^{pass})|^2 dz}$ \  
%where $\tau^{pass}$ is ``the lifetime'' of neutron passing through the slit with the upper absorber; it somehow depends on the slit width.
 
By virtue of our analysis, we consider the above ``principle for observation of the quantum gravitational states'' a pseudophysical (hence, uncriticizable) part of the authors' work.

\subsubsection{Discussion of the observation ( UCN transmission) method}
 
We also argue that conducted measurements, {\--} counts of transmitted neutrons by an external detector, in principle, cannot not reflect original quantum information, whatever it could be. First of all, ``the washing out'' effect is created because neutron paths from the last scattering points to the detecting points are quite randomly distributed over statistics of preceeding events occurring in a neutron wave guide. But the main point is that neutrons emerged from the slit are not confined, hence, not ``quantized'' anymore. Individual wave packets become rapidly spread over 3-space and in time. The measurements should be organized by some methods of neutron scanning inside the slit.

The raised questions cannot be properly addressed and clarified within the authors' {\em ad hoc} ''principle for observation''. Yet, our criticism cannot be complete without looking at the results of the experiment.

\subsection{What was observed?}

\subsubsection{The early and new observations}

There are several basic setups used in the experiment, such as: ``bottom mirror and upper rough scatterer or  (optionally) absorber'', ``two separated bottom mirrors at different elevations and upper rough scatterer or (optionally) absorber'', ``bottom mirror and two upper separated scatterers (one is fixed, the other's height  variable)''. Besides, neutron transmission measurements were conducted with the use of an integral detector and, sometimes, a PSD, {\--} a film containing heavy fissionable nuclei or (optionally) Boron-10 isotope. It is not clear from publications, which variants are methodologically more reliable, and which ones should be disregarded.

Readers, probably, are mostly familiar with early reports (2000-2002) on transmission measurements depicted in 
Fig. \ref{UCNtransmEARLY}. It looks well matching the stepwise picture required by the cited above ``principle for observation''. After comparing  Fig. \ref{UCNtransmEARLY}  with later, seemingly more accurate, results, we concluded that the advertised early results should be disregarded. See, for example, 
Fig. \ref{UCNtransmLATER},
 where ``stepwise'' pattern practically disappeared. Read the above cited statement  once again:{\em ``It was found, that except for the ground state, the stepwise increase is mostly washed out''}. Does it say that the data in Fig. \ref{UCNtransmEARLY} are mistaken? Indeed, the corresponding differential (PSD) measurements in 
Fig. \ref{Differential}
 do not indicate a resonance line picture. The authors themselves concluded from their data analysis that there is a great deficit of transmitted neutrons in the lowest state.

There are irregularities in integral and some PSD neutron distributions, which seem to be statistically not significant and hardly reproducible. As mentioned, they are, most likely, due to geometrical factors (diffraction on the exit window). 

%%%%%%%%%%%%%%%%%%%%%%%%%%%%%%%%%%%
\begin{figure}
\includegraphics{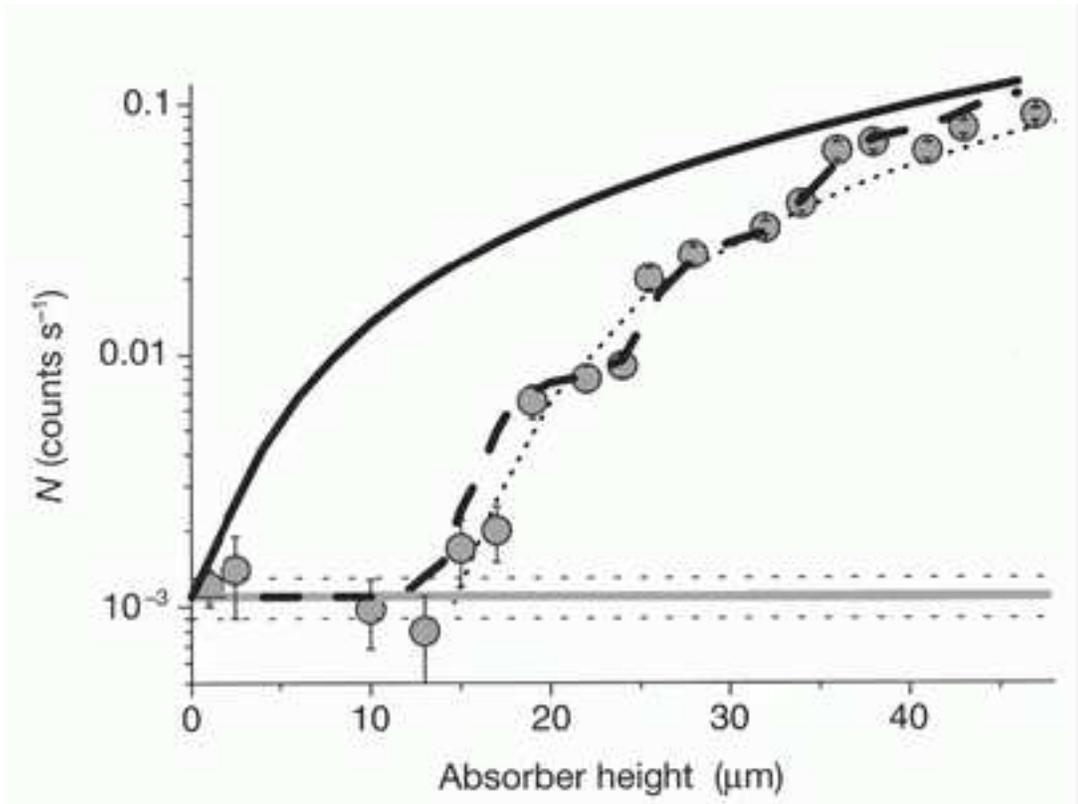}
\label{UCNtransmEARLY}
\caption{\label{UCNtransmEARLY} Early, apparently incorrect, integral measurements \cite{Nesvizh21}: neutron counts versus slit width. 
The circles with error bars show the experimental data. 
The solid line is the corresponding ``classical expectation'' model $N\sim a^{3/2}$. 
The dashed curve corresponds to the authors' QM phenomenological semi-classical treatment. 
The dotted curve is a model with no high energy state contributions. Notice differences between curves due to the absorption effect.}
\end{figure}

%Fig.\ref{UCNtransmEARLY}

\begin{figure}
\includegraphics{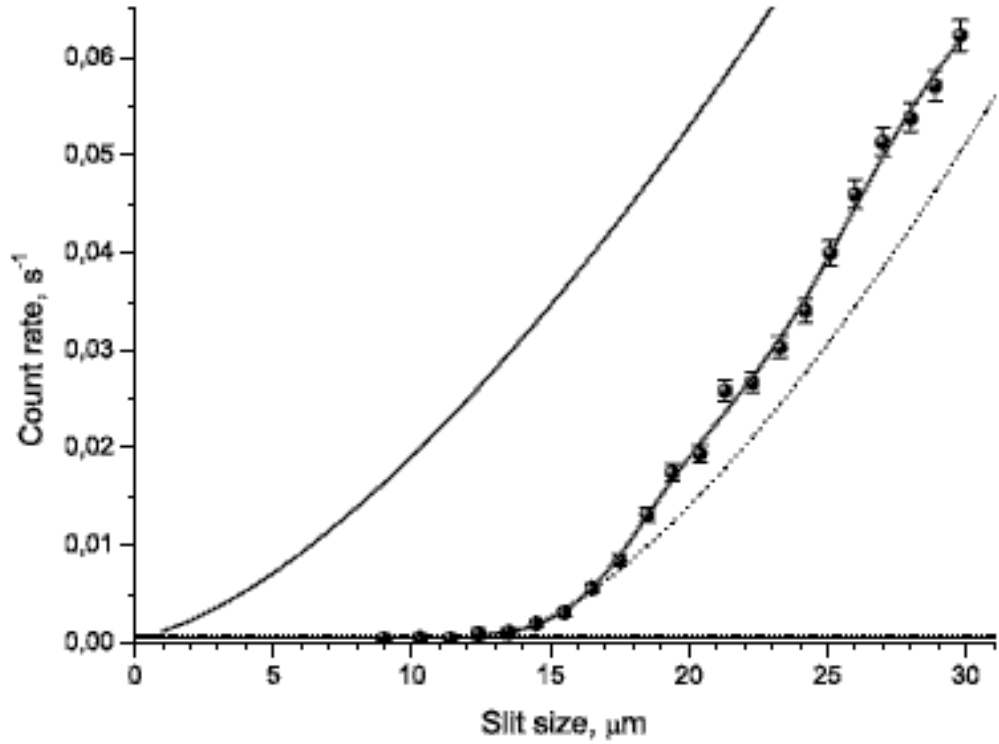}
\label{UCNtransmLATER}
\caption{\label{UCNtransmLATER} New measurements of neutron transmission: neutron counts versus rough scatterer/absorber height. 
The circles with error bars are the experimental data. The upper curve is a classical model $F_{cl}(a)\sim a^{3/2}$. The lower curve is a semi-classical model where only the lowest quantum level is taken into account. The experimental points are approximated by a model where higher levels are taken into account  \cite{Nesvizh10}, \cite{Nesvizh18} }
\end{figure}

\begin{figure}
\includegraphics{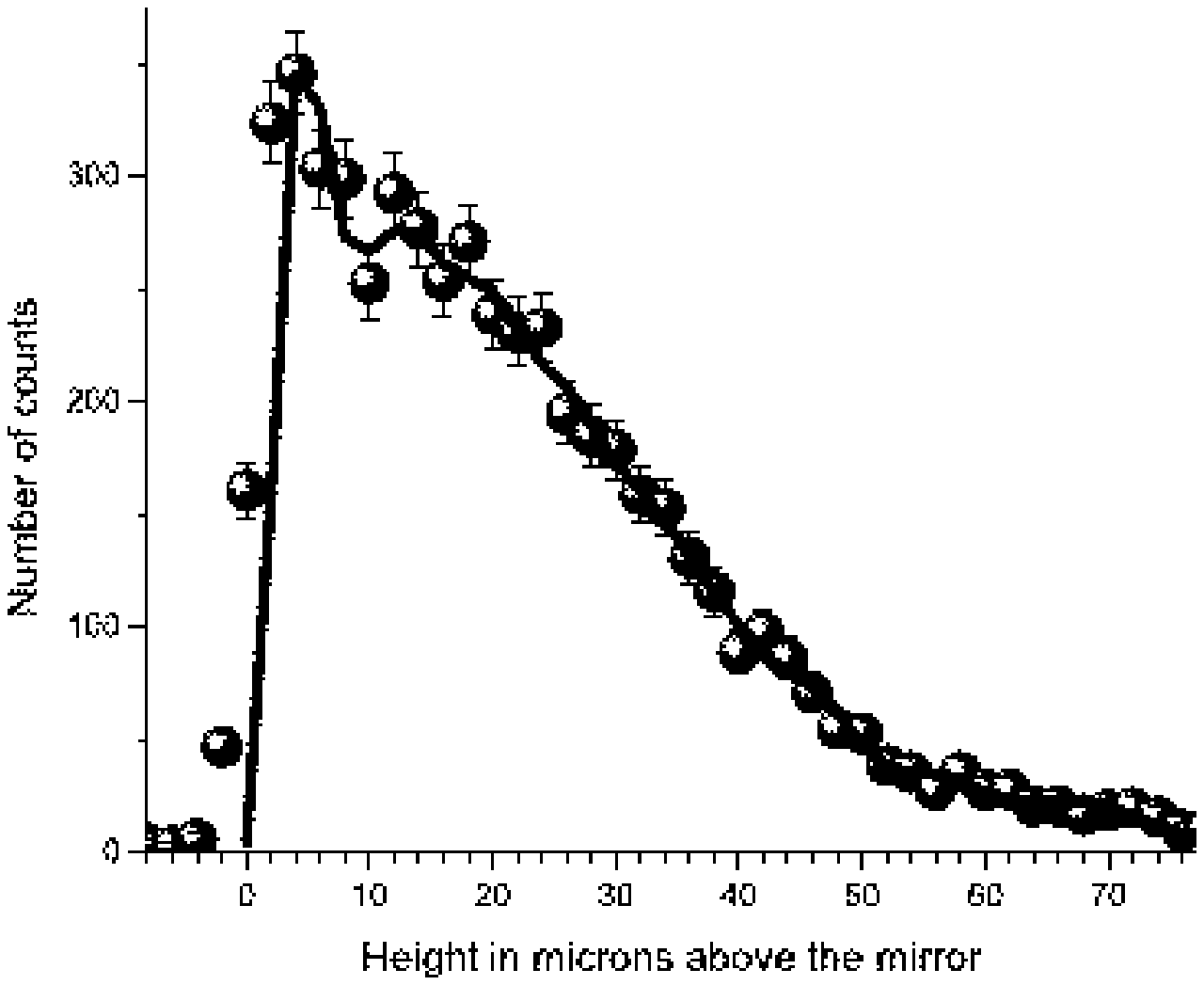}
\label{Differential}
\caption{\label{Differential} Differential neutron flux distribution $P(z)$ measured by a position sensitive detector (PSD): neutron count in $\Delta z$ interval for a fixed width $a$. The circles with error bars show the experimental data approximated by a phenomenological many-level model for ideal PSD spatial resolution. The width of slit $a$ is not indicated  \cite{Nesvizh18} }
\end{figure}

%%%%%%%%%%%%%%%%%%%%%%%%%%%%%%%%

\subsubsection{Summing up}

As of today, the authors claim that, at least, the lowest (ground) state of neutron in the slit has been observed and, possibly, next level resolved, see the cut-off in the transmission at absorber heights about 
$10 {\--} 15$\ $microns$, Fig.\ref{UCNtransmLATER}. The authors' treatment of the data is based on the semi-classical quantum model and  ``the principle of quantum state observation'' (stepwise integral transmission function). The fact that neutrons are not transmitted through the slit of small width (due to absorption) is considered by the authors as evidence of the observation of the first quantum state. But how one can admit that the absence of counts of perished in the slit neutrons is the evidence of presence of them in the ground state?

We have to reiterate the main results of our analysis, as next.  The neutron bouncer, as a QM object,  could not be formed in the real slit due to a number of reasons outlined in our analysis (first of all, neutron source incoherence).  The observed cut-off in the transmission function at small widths (less than about the maximal de Broglie wavelength $z_1 \sim 1.5\cdot 10^{-5}$\ $m$) is expected in the absorbing slit.  Anyway, the quantum bouncer could not be observed by the method of counts of transmitted neutrons with the use of an external detector because the neutrons which left the slit are not confined, hence, not quantized anymore. In other words, quantum states of neutrons in the gravitational field of Earth, indeed, have not been observed. 

We think that two textbook popular problems on energy quantization of confined particle (the particle in infinite well and the neutron bouncer) should be considered as toy model examples. Possibilities of their practical realization were never seriously discussed.  A neutron bouncer seems to be more complex QM object than that given by the 1D time-independent model. In particular, it should be considered within the concept of wavepacket spread in one bounce and over a series of bounces \cite{Donchenski}. Yet, a neutron in the slit on land is very much a macroscopic system, a neutron waveguide in a force field. So its theory must go well beyond the 1D Schrodinger-Airy mdel, with minimal phenomenology and semiclassical idealization and maximal rigorous Physics.

The key problems are how to create a quantum bouncer in a stable state and how to observe it. In the authors' physical methodology of the experiment, the existence of the quantum bouncer was taken for granted, and the straightforward objective was to observe it. This gave the authors a liberty to work with a semi-classical  phenomenological approach to experiment set-up and measurement treatment. We argue that the methodology was not adequate to complexity of the problem. In our view, the authors failed in a full theoretical comprehension of the problem and practical realization of the quantum bouncer concept. The very question about possibility of the realization remains open, first of all, because of the ambiguity of the concept. From the authors'  publications, one does not gain new insights into the problem.

We also criticize the authors' statements about a unique role of the Grenoble experiment methodology in fundamental physics studies, as next.

\subsection{Statements on a unique role of the Grenoble experiment methodology in fundamental physics studies}

\subsubsection{Methodological clams}

In our view, the methodology and the tone of the authors' publications is heavily influenced by wrong premises and claims of fundamental novelty and value of the Grenoble experiment. Methods of neutron wave optics were widely used for practical needs and in fundamental studies for decades. However, the authors specific statements that the UCN waveguide method in the Grenoble experiment opens new opportunities for studies in fundamental physics (see \cite{Nesvizh2} with further references) do not seem to have a real sense.

\subsubsection{Statement that universality of matter quantization was confirmed}

From \cite{Nesvizh22}: `{\em `The existence of quantum states of electrons in an electromagnetic field is responsible for the structure of atoms, and quantum states of nucleons in a strong nuclear field give rise to  
the structure of atomic nuclei. But the gravitational force is extremely weak compared to the electromagnetic and  nuclear force, so the observation of quantum states of matter in a gravitational field is extremely challenging...

Here we report experimental evidence for gravitational quantum bound states of neutrons...  Under such condition, the falling neutrons do not move continuously along the vertical direction, but rather jump from one height to another, as predicted by quantum theory. '' }

The standard QM description of ``gravitationally bounded'' non-radiating neutrons in free fall is given in terms of continuous de Broglie wave propagation with a varying wave number. The statement that ``falling neutrons do not move continuously along the vertical direction, but rather jump from one height to another'' is 
%primitively% 
just wrong.

\subsubsection{Spontaneous emission of graviton by a quantum bouncer}

The authors attempt to consider the (never observed) quadrupole gravitational radiation (a graviton) from a neutron bouncing in the slit  \cite{Nesvizh5}.  There are speculations in astrophysical and cosmological studies that gravitational waves can originate from aggregating and colliding super-dense high-mass objects such as black holes, binary stars and surrounding matter. The concept of graviton (a gravitational field quantum) comes from GR model of gravitational field in terms of linear perturbation of Minkowski space metric. The authors apply this concept for a neutron motion in the slit and used the suggested in literature quasi-classical formula for the quadrupole momentum of a QM system for the assessment of the radiation from state-to-state neutron transition in the slit. Not surprisingly, the assessed intensity of the radiation in the Grenoble experiment is completely vanishing. But why one should interested in this issue at all?

%The problem concerns Newtonian dynamics of point particle having a momentum $\vec p = m\vec v$ as a function of coordinates and time, in free fall in the gravitational field. In QM, the problem is treated in terms of the de Broglie wave concept $\lambda=h/p$ while a classical continuity of motion remains preserved (unless noticeable quantum transitions occur, what is not the case). Shouls it be ``jumps'' between quantum levels, one could think about validity of the EP. Thus, arguments on accepting or doubting the EP have principally equal validity in classical physics and QM. (We are not speaking about tentative physical models  where the EP is deliberately broken). 

\subsubsection{The Equivalence Principle verification in QM domain as the issue of QM foundations}

The Newtonian equation $m d^2 z/dt^2 =m g$ does not contain a particle mass, while the Schrodinger equation (\ref{1}) apparently does; its solution gives mass dependent wavefunctions with the scaling length $z_0= (h^2/2m^2 g)^{1/3}$. The authors' argument is that one can actually measure the neutron mass by observing the probability density for neutrons in the slit. Say, given $z_0$ measured, find 
 $m= (h^2/2 z_0^3 g)^{1/2}$. It looks like behavior of particles of various masses in free fall in QM is different from classical picture (the EP violation?). However, it is not true. 

The mass is canceled in the equation of motion or, equivalently, in energy conservation formula $2gz_0=v_1^2$, where $v_1=2 \bar v $ is a final speed of the accelerated motion ($\bar v$ is an average speed over distance $z_0$ passed). The matter is that the factor  $z_0$ expresses the equality $(h^2/2m^2 g)^{1/3}=h/p_0$\ ($p_0=mv_1$) based on the same assumption: the validity of gravitational conservation law (in terms of the Ehrenfest theorem) with the acknowledgment of the equality of gravitational and inertial mass (the EP). With this, the measured $z_0=h/p_0$ allows you to determine the only quantity, namely, $p_0=mv_1$. To find the mass, we need something else, for example, $v_1$ or $g$. The neutron mass comes out from $z_0$ exactly as in Newtonian mechanics. 

%In the following speculations by the authors, both $m$ and $g$ are supposed to be given. 

\subsubsection{Constraints on short-range forces}

In Modern Physics, hypothetical short-range forces are those rapidly decaying in comparison with the 
$1/r^2$ force. In other words, they act on a test particle significantly only near a source (for example, a Yukawa-type force. In 1984, the paper ``New macroscopic forces?'' by J. Moody in co-authorship with Nobel Laureate F. Wilczek was published, where new (short-range, very weak) hypothetical forces were suggested for testing \cite{Moody}. A typical test model is given in the form of potential combining the $1/r$ and Yukawa-type parts: 
\begin{equation}
\phi(r)=-(k/r)[1+ c \exp{-r/\lambda}]
\label{6}
\end{equation}
where $c$ is a relatively very small Yukawa-type contribution, and $\lambda$ is its effective length of action. Formally, it immediately takes the Newtonian form for $k=GM$, when $r>>\lambda $.

The work  \cite{Moody} was greatly motivated by extremely high sensitivity of modern Eotwash method to the perturbation of gravitational acceleration, about $10^{-14}$\ $m/s^2$ at that time \cite{Shortrange}. In the Grenoble experiment, the sensitivity to $g$ is extremely low (say, $\pm 25$\ $\%$ of $g$. It looks like there is nothing to be involved. Nevertheless, the authors claim that new constraints on the range of hypothetical forces can be put from the future Grenoble experiment  \cite {Nesvizh4, Nesvizh8,  Nesvizh11, Nesvizh12,  Nesvizh15, Nesvizh19, Critic}. The emphasis is made on a smallness of distance $r=z_0 \sim 10^{-5}$\ $m$ from the bottom mirror. 

The problem of the hypothetical force testing was formulated in  \cite{Moody} in terms of experiment sensitivity to $g$. Specifically, the assessment of force due to a semi-infinite axion source (the mirror in the Grenoble experiment)
was made. The corresponding acceleration exerted on a free neutron by an iron slab of thickness $\lambda\approx 10^{-5}$\ $m$ is about $10^{-12}$\ $m/s^2$. Therefore, the Eotwash method is good enough (compare with $\pm 25$\ $\%$ of $g$ in the Grenoble experiment).

%for example, in the perticle physics axion problem: the coupling of axion-like particles to matter. It is not clear why the authors make their own assessments of the effect range in comparison with the local Earth acceleration $g$, when such detailed assessments have been presented graphically in the original work \cite{Moody}. See, for example, the assessment of force due to a semi-infinite axion source. The corresponding acceleration exerted on a free neutron by an iron slab of thickness $\lambda\approx 10^{-5}$\ $m$ is about $10^{-12}$\ $m/s^2$. 

%In view of the above comments, the constraint issue is very simple. A sensitivity of the Grenoble experiment to the acceleration is $(dz_0/z_0)\sim {3\Delta g/g}$ while the hypothetical effect is of order  $\Delta g/g\sim 10^{-13}$. This means that in order  to observe the effect, the Grenoble experiment must be conducted at the fantastic precision level  $10^{-13}$ (the current accuracy having been about $.25$). 

Meanwhile, works in different physical laboratories are in progress in attempts to test hypothetical phenomena by observing their direct consequence followed from the corresponding field/particle theory. Physicists consider experiments aimed to answer the ``yes or no'' question rather than putting constraints. For example, in \cite{Axion}, a direct search for axion-induced force is conducted for milli-eV mass particles with axion-like coupling to two photons.

It seems that the authors in their claims put their UCN slit technique into perspective in connection with the GRANIT project \cite{Nesvizh2}. The plan is to study ``resonant modes'' in the slit via an interaction of magnetic moment with electromagnetic field with the hope of dramatically improvement of the precision. However, the above discussed problems of neutron injection would severely restrict resolution in the new project unless some new methodological ideas were suggested. 

It should be noted that an observation of neutron in the electromagnetic field in a resonant cavity  is a refrain of old QED studies, and, again, has nothing to do with quantum gravity.

\section{Conclusion}

Our detailed analysis of the Grenoble experiment led us to the conclusion that the experimental data do not support the claim that quantum states of neutrons in the gravitational field of Earth have been observed. The scientific value of the results and experiment methodology seems to be exaggerated in the authors' publications.

\end{document}